\documentclass{aastex}
\singlespace

\begin{document}

\title{Discovery of a Possible Anomalous X-ray Pulsar in the Small
Magellanic Cloud} \author{R.C.Lamb$^1$, D. W. Fox$^1$,
D.J. Macomb$^2$, T.A. Prince$^1$}

$^1$ Space Radiation Laboratory, California Institute of Technology,
Pasadena, CA 91125

$^2$ Physics Department, Boise State University, Boise, ID 83725-1570


\centerline{Accepted for publication in The Astrophyscial Journal Letters}

\begin{abstract}

We report the serendipitous detection of a previously unreported
pulsar from the direction of the Small Magellanic Cloud using data
from the {\it CHANDRA X-Ray Observatory}.  Because of its luminosity
($\sim 1.5\times 10^{35}$ ergs/s), its near lack of variability for
more than 20 years, and its very soft spectrum we propose that it is
an anomalous X-ray pulsar (AXP). Data from the {\it ROSAT} PSPC in
conjunction with the CHANDRA data give a period, $P$, of 5.44 s and a
spin down time, $P/\dot P$, of 11 ky.  If this is a correct
identification it will be the first extragalactic AXP and the fastest
yet discovered.

\end{abstract}

\section{INTRODUCTION}

Anomalous X-ray pulsars (AXPs) have a number of properties that
distinguish them as a class from other pulsars.  (See Mereghetti 2001
for a review.)  They have periods in the range 6-12s, X-ray
luminosities in the range $10^{34}-10^{36}$ ergs/s, very soft X-ray
spectra, little or no variability on time scales from hours to years,
and they undergo relatively steady spin-down with no evidence for binary
motion.  The limited number of members of this class ($\sim$ 6) has
inhibited development of a theoretical understanding of their
properties.  Models to account for their X-ray emission generally fall
into two categories depending on the energy source, either loss of
magnetic field energy (magnetar models) (Thompson \& Duncan 1996; Heyl
\& Hernquist 1997) or accretion.  Accretion models may be further
subdivided according to the source of infalling material; binary
companion models e.g. Mereghetti \& Stella (1995), or accretion from a
disk leftover from a supernova explosion (Marsden et al. 2001,
Francischelli \& Wijers 2002).

In this paper we present evidence for a possible addition to this
class.  In an examination of archival X-ray data from a number of
X-ray satellites (ROSAT, Chandra, RXTE) we have discovered evidence
for a previously unreported pulsar in the Small Magellanic Cloud.  The
properties of this pulsar are consistent with those of the AXPs.  By
virtue of the known distance to the SMC, 57 kpc (Feast \& Walker
1987), the luminosity can be rather accurately established and it
falls in the range of AXP luminosities.  If this is a correct
identification, then it will be the fastest yet discovered (P = 5.44
s) and the first extragalactic AXP if we discount the soft gamma
repeater, 0526-66, in the LMC.  This latter source exhibits all the
characteristics of an AXP when it is not bursting (Marsden et
al. 2001).

\section{OBSERVATIONS}

The discovery data were obtained from a 100ks ACIS-I observation which
began 2001 May 15, obsid 1881.  The position of the source is: 01$^h$
00$^m$ 43.14$^s$, -72$^\circ$ 11$'$ 33.8$''$ (J2000), approximately 10
arc minutes from the ACIS-I aim point.  The density of the stellar
field, and its location well out of the Galactic plane, mitigate
against ready boresight correction of the Chandra data.  Apart from
applying the prescribed aspect offset for this data set (See the
Chandra X-ray Center ``Aspect caveat'' page at
http://asc.harvard.edu/cal/ASPECT/aspect\_caveats.html), we have not
further refined the provided aspect solution.  Given the negligible
uncertainty in the source centroid ($\sim$0.05 arcsec), we expect an
absolute astrometric accuracy in line with the overall Chandra
performance to date, which has provided a 90\%-confidence radial
uncertainty of 0.6 arcsec; see
http://asc.harvard.edu/cal/ASPECT/celmon/index.html.

In accordance with the CHANDRA source naming convention, the source is
designated CXOU J0110043.1-721134.  A selection circle of 24 pixels
(11.8'') was chosen to encompass $>90\%$ of the source photons.  6099
photons were retrieved from the observation with a background
estimated to be 111 photons.  The times of arrival of the source
photons at the space craft were adjusted to the barycenter of the
solar system and a fast Fourier transform (fft) was performed.

Inasmuch as the ACIS-I CCD read-out time is a fixed 3.241 seconds, the
Nyquist limiting frequency for the fft is 1/6.482s = 0.154 Hz.  In
order to show clearly this limitation, the fft was performed with a
bin size of 0.1 seconds, and a portion of the resulting power spectrum
(normalized to unity power) is shown in figure 1.  There are three
peaks present, two of which are aliases of each other at 0.124706(1)
and 0.183828(1) Hz, on opposite sides of the Nyquist frequency.  These
peaks have highly significant power values of 38.4.  The chance
probability of such a power is $\sim 3\times10^{-17}$.  The peak at
0.308 Hz is due to the digitization limit (3.241 s) of the ACIS-I and
has a power value approximately equal to the number of photons.

\begin{figure}
\figurenum{1}
\epsscale{1.0}
\plotone{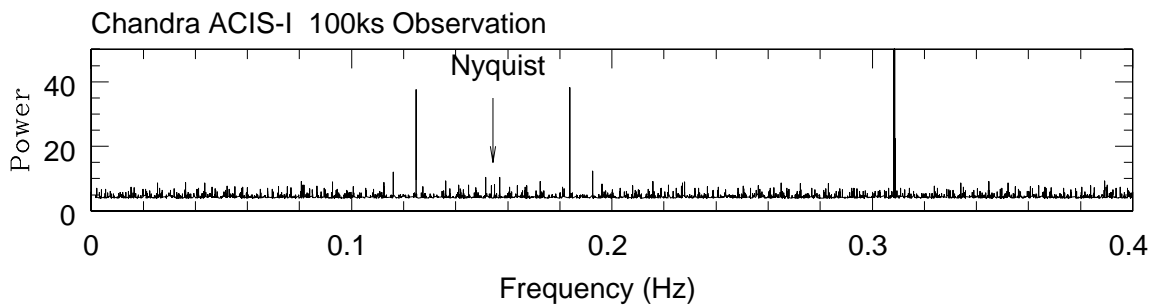}
\caption{FFT of CHANDRA Observation of CXOU J0110043.1-721134.  The
Nyquist frequency at 0.15427 Hz is indicated.  The two power peaks of
0.124706(1) and 0.183828(1) Hz are aliases of each other.  The
remaining peak at 1/3.241 = 0.3085 Hz is due to the limited time
resolution of the ACIS-I instrument.}
\end{figure}

The spectrum of the source is very soft.  An acceptable spectral fit
to the data is given by a black-body model with an absorption column
density 1.4$\pm 0.2\times 10^{21}$ H-atoms/cm$^2$ and a value of kT =
0.41$\pm$0.01 keV.  The fit to the data is shown in Figure 2.

\begin{figure}
\epsscale{0.5}
\figurenum{2} \plotone{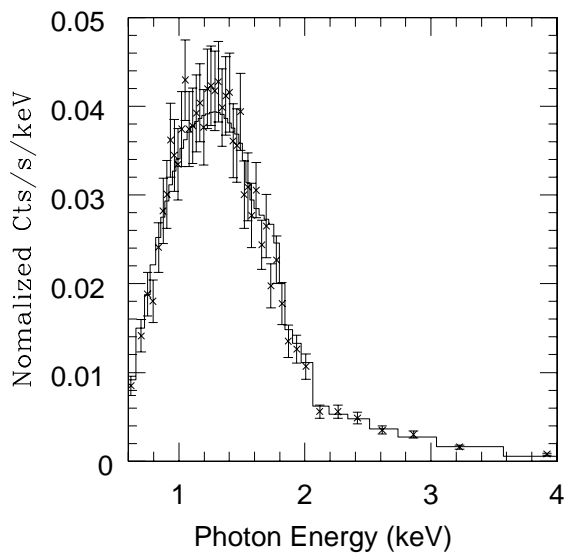}
\caption{Spectral Fit for CXOU J0110043.1-721134.  The fit assumes a
black-body model with an absorption column density 1.4$\pm
0.2\times 10^{21}$ H-atoms/cm$^2$ and a value of kT = 0.41$\pm$0.01
keV.  The fit has a reduced chi-squared value of 1.4 for 45 degrees of
freedom, and thus is an acceptable fit.}
\end{figure}

A search of imaging X-ray satellite archives for data for this source
has produced many observations dating from 1979.  These observations
are listed in Table 1. The most extensive source coverage is from the
ROSAT satellite.  For ROSAT we have restricted the observations shown
in the Table by excluding all observations which contain fewer than
$\sim$ 50 source photons.  This means that for the ROSAT HRI exposures
less than 7.5 ks are excluded.  For the PSPC exposures less than 1.5
ks are excluded as are observations in which the source is more than
30 arc minutes from the center of the field of view.  This latter
restriction is designed to avoid large vignetting corrections to the
counting rate.

\begin{deluxetable}{rrrrrrr}

\tablecolumns{7}

\tablecaption{List of Observations of CXOU J0110043.1-721134. The
luminosity is calculated for the interval 0.2-10 keV under the
assumption that the spectrum is that given in Figure 2}
\tablewidth{0pt}

\tablehead{\colhead{Satellite} & \colhead{Observ.}& \colhead{Date} &
\colhead{Exp.(ks)} & \colhead{Ct Rate (cts/s)} & \colhead{Mean Rate}
&\colhead{L (ergs/s)}}

\startdata

Einstein IPC  & 6297 & Apr 1980 & 23  & 0.0122(13) & & \\
& 3925 & Nov 1979 & 20 & 0.0144(17) & &\\
& & & & & 0.0133 (16) & 1.3$\times10^{35}$\\

ROSAT PSPC & rp500250n00 & Oct 1993 & 20. & 0.0275(16) & &\\
& 	rp600195a00 & Oct 1991 & 16. & 0.0273(14) & &\\
&       rp600195a01 & Apr 1992 & 9.2 & 0.0300(21)  & & \\
&       rp500142n00 & May 1993 & 4.8 & 0.0263(29) & &\\
&       rp600455a02 & Oct 1993 & 4.5 & 0.026(3) & &\\
&       rp600455a03 & May 1994 & 4.0 & 0.012(7) & &\\
&       rp600455n00 & Dec 1992 & 3.5 & 0.062(7) & &\\
&       rp600455a01 & Apr 1993 & 1.7 & 0.034(9) & &\\

& & & & & 0.0280(8) & 1.6$\times10^{35}$\\

ROSAT HRI & rh900445a01 & Apr 1995 & 34 & 0.0095(6)  & & \\
& rh900445n00 & Apr 1994 & 15 & 0.0105(9) & &   \\
& rh500137n00 & Apr 1993 & 14 & 0.0073(10) & &  \\
& rh500418a03 & Mar 1998 & 11 & 0.0071(10) & &  \\
& rh500418a01 & Oct 1995 & 8.2 & 0.0089(12) & &  \\
& rh500418a02 & May 1997 & 7.5 & 0.0098(13) & & \\
& & & & & 0.0090(8) & 1.5$\times10^{35}$\\

ASCA GIS & 55033000 & Nov 1997 & 72 & 0.016(4) &  & 1.8$\times10^{35}$\\

CH. ACIS-I & 1881 & May 2001 & 100 & 0.060(1) &  & 1.3$\times10^{35}$\\

\enddata

\end{deluxetable}

The luminosity values given in the table are derived under the
assumption that the source is associated with the SMC (distance 57kpc,
Feast \& Walker 1987) and that the source spectrum has the same black-body
model form as determined by the fit to the CHANDRA data.  The luminosity
is calculated for the CHANDRA spectral range 0.2 to 10 keV.

From the Table we notice a near constancy of flux values for a given
instrument.  The two Einstein observations are consistent.  The HRI
fluxes are consistent with a constant flux (chi-squared probability
6\%).  There is a single high PSPC observation (rp600455n00) which is
nearly 5 $\sigma$ above the mean.  Without that observation the
remaining PSPC count rates are consistent with a constant flux
(chi-squared probability 26\%)

The luminosity values from different detectors are also nearly
consistent with one another, averaging to a value of
1.5$\times10^{35}$ ergs/s.  ROSAT observations which are not
included in the table (e.g. PSPC observations with the source further
than 30 arc minutes from the center of the field) show the source as
well.  Thus the the variability of this source on a time scale of
months to years is relatively small.  In addition, within the 100ks
CHANDRA observation there is no significant variation of the source
count rate.

We may use the CHANDRA data as a guide to calculate what fft power
values are to be expected from the other observations.  The limited
time resolution of the ACIS-I data and the fact that the pulsed signal
is near the Nyquist limiting frequency of 0.154 Hz prohibit any
detection of harmonic content higher than the fundamental.  We
therefore make the assumption that the pulsations are purely sinusoid,
in which case the power expected is 0.25 $N_{pulsed}^2/N_{tot}$, where
$N_{pulsed}$ is the number of pulsed photons and $N_{tot}$ is the
total number of photons.  By setting the power to be the observed
value of 38 and taking into account the 2\% background, we find that
the pulsed fraction is 16$\pm$3\%.

As a further attempt to detect pulsations from CXOU J0110043.1-721134
we analyzed data from the observation (rp600195a00) which has the
highest sensitivity to a pulsed signal.  It has the second longest
exposure (16 ks) of any of the PSPC observations and has by far the
shortest duration (85 ks).  Other comparable exposure PSPC and HRI
observations have durations more than 15 times longer and their
sensitivity to a periodic signal is diluted by the search range needed
to cover potential frequency variation over this length of time.  The
duration of rp600195a00 is such that no phase slippage is expected for
$\dot P$s in the range of the AXPs.

The source for this observation is within 12.7' of the center of the
PSPC field of view where the angular resolution is excellent, thus a
selection circle of 80 pixels (40'') could be used.  With this
selection 392 photons were retrieved, of which an estimated 37.6$\pm1$
are background.  A comparison of the pulse height spectrum for the
source and the background, showed that sensitivity to a pulsed signal
could be enhanced by eliminating photons with energies less than 0.4
keV.  With this restriction 358 photons remained.  The times of
arrival of these photons were adjusted to the barycenter of the solar
system and an fft was performed.

We restricted our search region to be near one of the two possible
frequencies seen in the CHANDRA observation, either 0.1247 or 0.1838
(Hz).  To account for a frequency change over the 9.6 years between
the two observations we assume that the object has characteristics of
the known AXPs.  This implies that the pulsar is spinning down, and
that the time scale for this, $P/\dot P$, is greater than 6.8 ky (see
the table in Mereghetti 2001).  In the search we have allowed for a
possible time scale as short as 1 ky.

With these assumptions the range of frequencies to be searched is 1.20
mHz at the lower frequency and 1.76 mHz at the upper frequency.  The
ffts near these frequencies for both the CHANDRA observation (a) and
(b) and the PSPC observation are shown in Figure 3.  In the lower
frequency search region there is no significant power peak, however in
the upper region there is a peak power of 15.4 which may be
significant.

To assess its significance we use a formula first derived by Fisher
(1929).  This formula gives the probability that a given Fourier
power, P, taken from a range of n independent values of power, will be
exceeded by chance.  Equation (4) of Fisher gives that probability as
a series, whose leading term, $n(1-g)^{n-1}$ is the only term which is
significant for our values of P and n.  The parameter ``$g$'' is the
fraction of the power contained in the term in question.  We have used
normalized values of power, thus $g = P/n$.  However, since we have
used a frequency digitization finer than the independent frequency
spacing, 1/T, we must multiply Fisher's expression by an oversampling
factor.  For the fft's we have used a frequency spacing of $\sim$ 1/5T
for which an appropriate value of the oversampling factor is 3 (Lewis,
1994).

Putting in the numbers we find a chance probability of $1.0\times
10^{-4}$ of finding such a power of 15.4 or greater within either
search range.  We judge this to be sufficiently small to be a
detection.  Using the PSPC observation which occurred 3705 days
prior to the CHANDRA observation, we derive a frequency derivative
of $-5.08\pm 0.07\times10^{-13}$ Hz/s.  This corresponds to a time
scale, $P/\dot P$, of 11 ky.

In addition to the peak at 0.183982(2) Hz there are two other peaks
in its vicinity with powers greater than 8.  They are separated from
the main peak by 0.000173 Hz and 0.000519 Hz.  These peaks are due to
the on-time profile (window function) of the PSPC data.  The 0.000173
Hz frequency is a beat frequency with the orbital period of the
satellite (96 minutes); the 0.00519 Hz frequency is its third
harmonic.  In a simulation of the dataset we have reproduced this
behavior.

\begin{figure}
\epsscale{1.0}
\figurenum{3}
\plotone{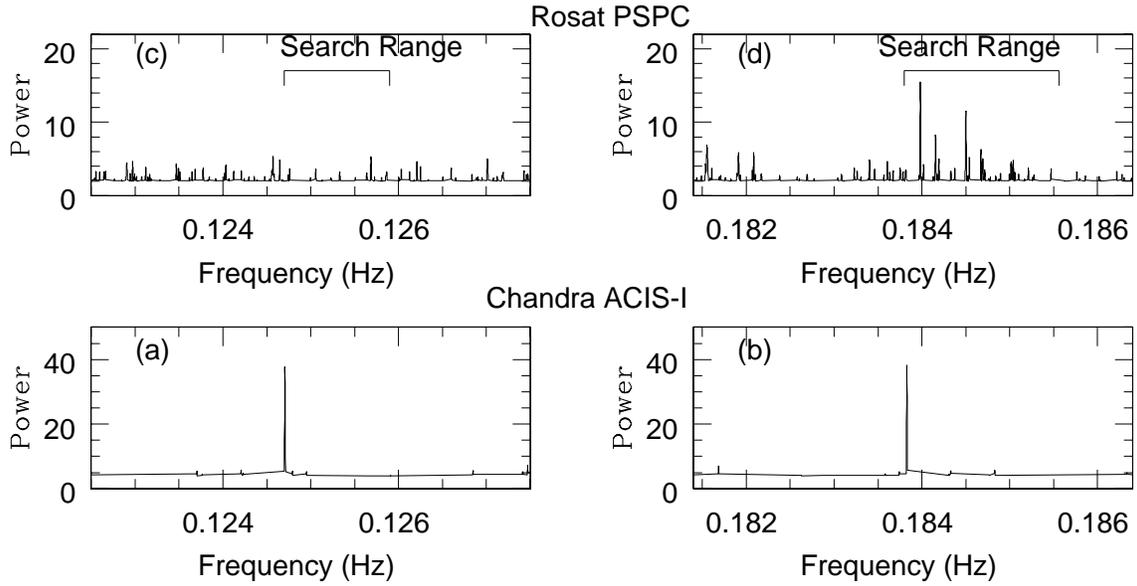}

\caption{FFT of CHANDRA (a) and (b) and ROSAT PSPC Observations (c)
and (d) in the vicinity of 0.125 and 0.184 Hz.  In (c) and (d) the
search ranges are indicated.  The peak is (d) is significant with a
chance probability of $1.0\times 10^{-4}$ (see text for details),
indicating that this is the right frequency for CXOU
J0110043.1-721134.  The two other nearby peaks in (d) at power values
of 8 and 11 are due to the on-time profile (window function) of the
dataset.}
\end{figure}

The pulse profile of the PSPC observation is shown in Figure 4.  There
is no evidence in the fft for harmonics higher than the fundamental.
The pulsed fraction is 36$\pm$5\%.  This disagrees with the 16\% value
derived from the CHANDRA detection.  This discrepancy can be accounted
for by the different spectral response of the two detectors and a
difference between the unpulsed and pulsed spectrum.  Relative to the
ACIS-I detector the PSPC weights lower energy photons significantly
more than higher energy photons.  For example, the ratio of effective
area for the PSPC at 0.8 keV to its effective area at 2.0 keV is
$\sim$ 3.  The same ratio for the ACIS-I detector is $\sim$0.8.
Further we find that the pulsed source spectrum is somewhat softer
than its unpulsed emission.  Therefore because of the PSPC's energy
response it will detect the pulsed part of source emission more
efficiently than the unpulsed part.  This qualitatively accounts for
the higher pulsed fraction as determined by the PSPC.

We have calculated the sensitivity of the remaining non-CHANDRA
observations using this value of the pulsed fraction.  We find that
none of the other non-CHANDRA observations listed in Table 1 would be
sensitive to this pulsation.

\begin{figure}
\epsscale{0.5}
\figurenum{4}

\plotone{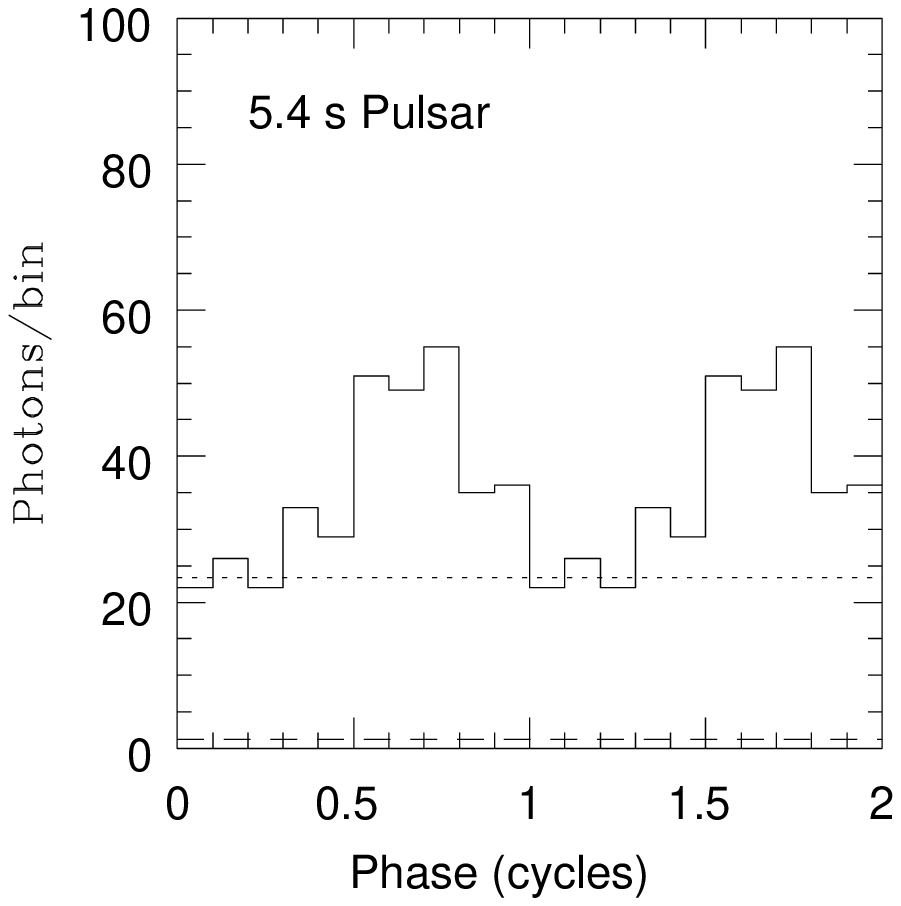}
\caption{Pulse Profile of PSPC Observation (rp600195a00 see table 1).
The lower dashed line indicates the instrumental background levels.
The upper dashed line gives the level of the unpulsed source
emission.}
\end{figure}

We have searched for a possible radio/optical counterpart to CXOU
J0110043.1-721134 and find no radio counterpart.  There is an 18th
magnitude star (S01020208968 in the Guide Star Catalogue-II available
on-line through http://vizier.u-strasbg.fr/viz-bin/cat?I/271\#sRM99.3)
which is located 1.2 arc seconds from the source.  The position of
CXOU J0110043.1-721134 has an uncertainly of 0.6 arc seconds,
therefore it is unlikely to be associated with this star.
Nevertheless it remains a possibility pending refinement of the X-ray
source position.  From the density of stars in the Guide Star
Catalogue within 30 arc seconds of this position, there is a 4\%
probability of a chance association within 1.2 arc second.  Multi-band
optical photometry has been performed on this star by Naz\'e et al
(2002).  The authors believe this star may be an early B star.  If
this star is a binary companion of CXOU J0110043.1-721134 then it
would be the first seen for an AXP and would raise the possibility of
accretion as the source of its luminosity.

\section{DISCUSSION}

The properties exhibited by CXOU J0110042.8-721132 are fully
consistent with the AXP class of X-ray pulsar (Mereghetti 2001 and
references therein).  Its spectrum is very soft - consistent with other
AXPs.  Its luminosity, $\sim 1.5\times 10^{35}$ ergs/s/cm$^2$, is in
the range of AXP luminosities ($10^{34} - 10^{36}$ ergs/s/cm$^2$).
There is little variation in its long-term and and short-term
intensity, and it is spinning down with a characteristic age, $P/\dot
P$, of 11 ky, typical of 3 of the 6 AXPs.

One of the central issues for AXPs is their energy source.  If
accretion then there may be a companion from which the accretion
occurs and which will cause a modulation in the pulsation unless the
orbit is observed face on.  (Some models of accretion powered AXPs
derive their accretion from a remnant disk leftover from a supernova
explosion (Marsden et al. 2001, Francischelli \& Wijers 2002).)  The
100ks continuous CHANDRA observation provides an opportunity for a
search for binary induced modulation.

We have subdivided the observation into four 25 ks segments, and
separately performed an fft on each.  The power peak in the 0.184 Hz
region shows no significant frequency variation.  In addition, we have
explored what would happen to the time series of photons from the
source if a hypothetical binary modulation were superimposed.  For
binary orbits with periods less than one day we find that a projected
semi-major axis of more than 0.5 light seconds will reduce the power
of the signal by more than one standard deviation in the power.  We
take this to be an effective limit on the size of any orbit.  For
orbital periods longer than the 1.2 day observation this size limit is
weakened.

This absence of any apparent binary modulation on a time scale $\le$ 1
day is consistent with evidence for other of the AXPs (Mereghetti
2001) for the lack of a binary companion.
	
Three of the AXPs appear to be associated with relatively young ($<$
20 ky) supernova remnants.  In each of these cases: (1E 2259+586: Rho
\& Petre 1997, Parmar et al. 1998); 1E 1841-045, Helfand et al. 1994;
and J1844-0288, Vasisht et al. 2000) there is evidence for extended
X-ray emission.  We have examined the CHANDRA data for evidence of
emission beyond that which is consistent with a point source.  We find
none.  However, a better limit on any possible extension to the source
may be derived from ROSAT HRI observations (the first two in the
Table) in which the source was approximately 6 arc minutes from the
center of the field where the angular resolution of the HRI is better
than in CHANDRA data 10 arc minutes off-axis.  One-dimensional
profiles though the image in the right ascension and declination
directions of the combined HRI dataset are consistent with a point
source with a width ($\sigma$) of 2.0 arc seconds.  There is no
evidence for emission at a level $>$ 5\% the counting rate at the peak
of the profiles at a distance of 10 to 20 arc seconds from the source.
Any extension to the source must be a distance from the source less
than the width of the point spread function of the HRI, 10 arc
seconds.  This corresponds to a limit to the size of any extended
X-ray emission of 2.8 pc at the distance of the SMC.

This limit is comparable to the X-ray extension observed for the two
SNR/AXP associations with known distances (Kes 73/1E 1841-045 and
CT109/1E 2259.1+586).  A deep CHANDRA exposure with the source
centered in the field of the HRC is needed to constrain strongly the
possibility of a supernova remnant association for this pulsar.

\acknowledgements

This work made use of software and data provided by the High-Energy
Astrophysics Archival Research Center (HEASARC) located at Goddard
Space Flight Center.

\end{document}